\documentclass[12pt,preprint]{aastex}

\def\ros{{\sl ROSAT}}
\def\chan{{\sl Chandra}}
\def\asca{{\sl ASCA}}

\def\ns{1E~1207.4--5209}

\begin{document}

\title{
Discovery of absorption features in the X-ray spectrum
of an isolated neutron star}

\author{
D. Sanwal\footnote{The Pennsylvania State University,
525 Davey Lab, University Park, PA 16802, USA},
G.~G.~Pavlov$^{1}$,
V.~E.~Zavlin\footnote{Max-Planck-Institut f\"ur Extraterrestrische Physik,
D-85748 Garching, Germany},
and M.~A.~Teter$^{1}$
}
\begin{abstract}
We observed 1E 1207.4--5209, a neutron star in the center of the supernova
remnant PKS 1209--51/52, 
with the ACIS detector aboard the {\sl Chandra} X-ray observatory and
detected two absorption features in the source spectrum. The features are
centered near 0.7 keV and 1.4 keV, their equivalent widths are about 0.1 keV.
We discuss various possible interpretations of the absorption features
and exclude some of them.
A likely interpretation is that
the features are associated with atomic transitions
of
once-ionized helium in the neutron star atmosphere
with a strong magnetic field. 
The first clear detection of absorption
features in the spectrum of an isolated neutron star provides an
opportunity to measure the mass-to-radius ratio and constrain the
equation of state of the superdense matter. 

\end{abstract}


\section{Introduction}

Although neutron stars (NSs) have been studied
extensively  for more than three decades, the
properties of the superdense matter in their interiors
still remain an enigma.  We know neither
the density nor the composition of a NS core.
We are not even sure that NSs are indeed composed of
neutrons --- for instance,
their cores could be pion or kaon condensates or a quark-gluon plasma.
A key observational property that would help  understand the true nature
of these objects is the mass-radius relation --- if we knew the
masses and radii for a sample of NSs, we could  compare them with the
predictions of models based on various equations of state of the
superdense matter, which  are quite different for NSs of different
composition
(see Lattimer \& Prakash 2001 for details).
A useful constraint can be obtained via measuring the
gravitational redshift $z$ of lines in the spectrum of thermal
radiation emitted from the NS surface (atmosphere), which directly
gives the mass-to-radius ratio:  $M/R = (c^2/2G) [1-(1+z)^{-2}]$.
However, many attempts to detect spectral lines
in thermal radiation from isolated (non-accreting) NSs
have been unsuccessful. For example, no spectral lines have been found
in the spectra of the Vela pulsar (Pavlov et al.\ 2001),
anomalous X-ray pulsar 4U~0142+61 (Juett et al.\ 2002),
and nearby radio-quiet NS RX~J1865--3754 (Burwitz et al.\ 2001;
Drake et al.\ 2002),
despite sensitive observations with 
the {\sl Chandra} grating spectrometers.
In this Letter we report the first firm detection of absorption
features in the spectrum of an isolated neutron star, \ns.

\ns, a radio-quiet central source of the SNR PKS~1209--51/52 
(also known as G296.5+10.0), was discovered by Helfand \& Becker (1984)
with the {\sl Einstein} observatory. Mereghetti, Bignami \& Caraveo (1996)
and Vasisht et al.~(1997)  interpreted the \ros\ and \asca\ spectra of
\ns\ as blackbody (BB) emission of $T\simeq 3$~MK from an area with radius
$R\simeq 1.5\, (d/2\, {\rm kpc})$~km. 
Zavlin, Pavlov \& Tr\"umper (1998) interpreted
the observed spectra as emitted from a light-element (hydrogen or helium)
atmosphere.  For a NS of mass $1.4~M_\odot$ and radius 10~km, they
obtained a NS surface temperature $T_{\rm eff}=(1.4$--$1.9)$~MK and a
distance $d=1.6$--3.3 kpc, consistent with the distance to the SNR,
$d=2.1^{+1.8}_{-0.8}$~kpc 
from the neutral hydrogen absorption measurements
(Giacani et al.\ 2000). 

Zavlin et al.\ (2000) observed
\ns\ with the \chan\ X-ray Observatory 
and discovered a period of about 424 ms, which proved that the
source is a NS. Second \chan\ observation provided an estimate of the
period derivative,  $\dot{P} \sim (0.7$--$3) \times 10^{-14}$ s s$^{-1}$
(Pavlov et al.\ 2002a).  This estimate implies that the characteristic age
of the NS, $\tau_c \sim 200$--1600 kyr, is much larger than the 3--20 kyr age
of the SNR (Roger et al.\ 1998), while the conventional magnetic field,
$B\equiv 3.2\times 10^{19} (P \dot{P})^{1/2}\, {\rm G} = (2$--$4)
\times 10^{12}$  G, is typical for a radio pulsar.  Spectral analysis of
these \chan\ observations, which resulted in the discovery of two
absorption features in the NS spectrum, is presented below.

\section{Observation and Data Analysis }

\chan\ observed \ns\ on 2000 January 6-7 and 2002 January 5-6 with the
spectroscopic array of the Advanced CCD Imaging Spectrometer (ACIS) in the
Continuous Clocking (CC) mode. This mode provides the highest time resolution
of 2.85~ms available with ACIS by means of sacrificing spatial resolution in
one dimension.  The source was imaged on the back-illuminated chip ACIS-S3.
The effective exposure times of the two observations were 29.3 ks and 31.6 ks,
detector temperatures $-110\, {\rm C}$ and $-120\, {\rm C}$, respectively.
In both observations the source-plus-background spectrum was extracted from
a segment of $4''$ length centered at the source position; the
background was taken from $20''$ 
segments in the one-dimensional
images. The source count rates after background subtraction are
$0.76\pm0.01$~s$^{-1}$ and $0.64\pm0.01$~s$^{-1}$ in the first and second
observations, respectively.
The background contributes only about $3\%$ of total counts, in the 0.4--5 keV
energy band.
The lower count rate in the second observation
can be attributed to the reduction of the ACIS sensitivity at low energies
with time, apparently caused by increasing deposition of 
a contaminant onto the ACIS filter\footnote{See the CXC web-page 
{\tt http://asc.harvard.edu/cal/Links/Acis/acis/Cal\_projects/index.html}}.

The continuum models (e.g. BB, power-law, fully ionized NS atmosphere) 
failed to fit the source spectrum, which deviates very
significantly from any of the models in a 0.5--2.0 keV  range,
showing two absorption features near
0.7 keV and 1.4 keV.
Another absorption feature observed near 2 keV
remains uncertain due to the calibration uncertainty near the  Ir~M
line 
from the telescope mirror coating,
so we do not consider this feature further.  We 
verified that the features at 0.7 and 1.4 keV are not due to
uncertainties in the responses and background subtraction and
concluded that they are intrinsic to the source.  We also checked
that no spectral features at these energies are seen
in  other \chan\ observations with the same observational setup
(e.~g., observations of PSR B1055--52 
and B0656+14).

To characterize the absorption features, we use
the spectral range from 0.4 to 2.5 keV
and model the continuum as an absorbed BB
with a temperature of $kT=0.26$~keV.  
Since we do not know the exact nature of the observed features,
we use three phenomenological models to check how sensitive
the feature parameters are to the choice of model:\newline
lines produced by an absorbing layer with Gaussian profiles of 
absorption coefficients:
\begin{equation}
F(E) = F_c(E)\, \prod_{i=1,2} \exp\left\{-\tau_i\,\exp\left[-\frac{(E-E_i)^2}{2\Gamma_i^2}\right]\right\}~,
\end{equation}
\noindent
multiplicative absorption lines with Gaussian profiles:
\begin{equation}
F(E) = F_c(E)\, \prod_{i=1,2}  \left\{1- r_i\,\exp\left[-\frac{(E-E_i)^2}{2\Gamma_i^2}\right]\right\}~,
\end{equation}
\noindent
lines with Gaussian profiles subtracted from the continuum:
\begin{equation}
F(E) = F_c(E) - \sum_{i=1,2} F_c(E_i)\, r_i
\exp\left[-\frac{(E-E_i)^2}{2\Gamma_i^2}\right]~.
\end{equation}
In  these models,
$F_c(E)$ is the continuum spectrum, $E_i$ and $\Gamma_i$ are the central
energy and Gaussian width of the $i$-th line, $\tau_i$ and $r_i$
are the parameters characterizing the line depth
($\tau_i$ is the optical depth at the line center, and
 $r_i=1 - F(E_i)/F_c(E_i)$ 
is the relative line depth, when the lines do not overlap,
$|E_2-E_1|\gg \Gamma_i$).

An example of the fit to the January 2000 data with 
model (1) is demonstrated in the upper panel of Figure~1.
The lower panel shows the continuum model
from the upper panel and the residuals to demonstrate the profiles of
the two absorption features.
Including the lines in the spectral model improves the quality of the
fit substantially: $\chi_\nu^2 =1.05$ for 232 degrees of freedom [d.o.f.]
versus $\chi_\nu^2 =2.5$ for 238 d.o.f.
We find similar improvement in the fit quality
for each of the line models for both observations.

The fitting parameters for models (1)--(3) are presented in Table 1.
The continuum normalization, $R=(1.64\pm 0.02) (d/2.1\,{\rm kpc})$ km,
is essentially the same for all of the models for both observations.
On the contrary, the hydrogen column density $n_{\rm H}$ is
substantially larger in the second observation. We interpret this as
the result of an increased contamination
of ACIS in the second observation which reduces the effective
area at lower energies$^3$ and can be crudely modeled as an additional
interstellar absorption (i.e., the $n_{\rm H}$ parameter does not
represent the true interstellar absorption). 

The line parameters for models (1) and (2) are very close to each other
within one observation. Model (3) results in somewhat lower central
energies.  The equivalent widths of the lines, calculated as 
$W=\int_{E_{\rm min}}^{E_{\rm max}} [1-F(E)/F_c(E)]\,\, {\rm d}E$
($E_{\rm min}=0.5$ and 1.1 keV, $E_{\rm max}=1.1$ and 2.5 keV,
for the 0.7 keV and 1.4 keV lines, respectively), are virtually
the same for the three models in a given observation.

For a given model, the parameters of the 1.4 keV line are almost
the same (within $1.2\sigma$) for the
two observations. The 0.7 keV line shows an apparent decrease
of the central energy, width, and equivalent width.
We attribute this to the effect of CCD contamination$^3$ that
results in an artificially large $n_{\rm H}$ and distorts the
parameters of the line. Since the contamination was presumably
insignificant in January 2000, we believe that the parameters
inferred from the first observation give more adequate description
of the lines.

For the empirical line models used, the best-fit models show
residuals 
suggestive of asymmetric and/or multiple lines. Since
the spectral resolution of the detector, about 100 eV at these
energies, is comparable to the model line widths $\Gamma_i$, each
of the observed features can be fit equally well with 
multiple line models, with smaller intrinsic line widths.
The observed broad features could also be produced 
due to a nonuniform magnetic field on the surface of the
rotating NS.

Reanalysis of the previous X-ray observations of this source has shown
no features 
in the \ros/PSPC and \asca/GIS spectra, which can
be readily explained by the very poor spectral resolution of these
proportional counters. The spectrum acquired with the \asca\ SIS (a CCD
imaging spectrometer) does show features at the same energies,  albeit
at low significance due to 
insufficient statistics.  The spectral models with
two lines, which fit the \chan\ spectra, are consistent with the
\ros\ and \asca\  spectra.

\section{Discussion}

Attempts to explain the absorption features
as caused by the intervening interstellar
(or circumstellar) material lead to huge overabundance for some elements.
For instance, the  absorption feature at 1.4 keV might be interpreted
as a magnesium photo-ionization edge at $E=1.305$ keV.
The fit of this feature with a photo-ionization edge model gives the
best-fit edge energy of 1.29 keV, not strongly  different from the
Mg K edge. However, to explain the strength of absorption, this model
requires the Mg abundance about 600 times the solar value. The energy of
the other feature, 0.7 keV, is close to the F K$_\alpha$ (0.67 keV) and
Mn L$_\alpha$ (0.64 keV) energies, but even larger overabundance of
these elements  would be needed to explain the observed 
feature.
Therefore, the absorption in the ISM looks improbable, and  the
observed features are most likely  intrinsic to the NS.

There are two potential ways for these absorption features to be generated
in the NS atmosphere --- cyclotron lines and atomic transition lines.

\subsection{Cyclotron lines}

First we consider that the observed features are cyclotron lines produced
in a strongly ionized NS atmosphere. If one assumes these are electron
cyclotron lines,  the features could be interpreted as the fundamental 
and the first harmonic of the electron cyclotron energy $E_{ce}=1.16 B_{11}$
keV in a  magnetic
field $B_{11}\equiv B/(10^{11}\, {\rm G})\sim 0.6\, (1+z)$
or as two fundamentals emitted from two regions with different magnetic
fields, $B_{11}=0.6\,(1+z)$ and $1.2\,(1+z)$, broadened by the
radiative transfer effects and/or nonuniformity of the magnetic field.
However, it is difficult to reconcile this interpretation with the 
expected strength of the surface magnetic field.
The measured $P$ and $\dot{P}$ imply a magnetic moment 
$\mu \sim (3 I c^2 P\dot{P}/8\pi^2)^{1/2}\sim 3\times 10^{30} I_{45}^{1/2}$
G cm$^3$,  where $I=10^{45} I_{45}$ g cm$^2$ is the moment of inertia.
This magnetic moment corresponds to a magnetic field
$B_e\sim 3\times 10^{12}$ G at the magnetic equator, if the field is a
centered dipole and the NS radius is $R=10$ km.  If the dipole is
off-centered, or there are substantial multipole components, the maximum
values of the surface magnetic field would be even higher. Although the
inferred  value of $B_e$ is based on $\dot{P}$ estimated from two
observations under the assumption of  a uniform slowdown (no glitches, no
timing noise, the NS is not in a binary orbit),  the discrepancy of the
fields looks uncomfortably large. In addition, the oscillator strength
of the first harmonic is  smaller than that of the fundamental by a factor
of $\sim E_{ce}/(m_ec^2) \sim 2\times 10^{-3}$ (at $E_{ce}\gg kT$
--- Pavlov, Shibanov, \& Yakovlev 1980),
so that it is hard to explain why the 1.4 keV feature is as strong as
the 0.7 keV feature if we assume the two lines are associated with the
same magnetic field.

The magnetic fields of $\sim 10^{11}$ G should exist within the NS
magnetosphere, and the resonant cyclotron scattering in such magnetic
fields could absorb (scatter) $\sim 1$ keV photons emitted from the NS
surface (e.g., Rajagopal \& Romani 1997).  In a dipole magnetic field,
$B(r)=B_e\, (1+3 \cos^2\theta_B)^{1/2}\, (R/r)^3$, the resonant scattering
of photons with energy $E$ occurs at a distance from the NS
surface $r\approx 3.7\, R\,
(B_e/3\times 10^{12}\, {\rm G})^{1/3}\, (E/0.7\, {\rm keV})^{-1/3}\,
(1+3 \cos^2\theta_B)^{1/6}$, where $\theta_B$ is the magnetic colatitude.
To produce a line with a width $\delta E$, the thickness of the scattering
layer should be $\delta r\sim (1/3) (\delta E/E)\, r$, i.e.,
$\delta r \la 0.03\, r \sim 1$ km for the observed $\delta E/E\la 0.1$.
It is not clear how two narrow layers (an analog of the Van Allen
radiation belts in the Earth's magnetosphere?) could be formed. Moreover,
the electron number density needed to obtain a sufficient optical
thickness $\tau$, $n_e\sim 10^{13}\, (\tau/0.5)\, (R/10\,
{\rm km})^{-1}\, (E/0.7\, {\rm keV})^{4/3}\,
(B_e/3\times 10^{12}\, {\rm G})^{-1/3}\, (1+3 \cos^2\theta_B)^{-1/6}$ cm$^{-3}$,
significantly exceeds the Goldreich-Julian density,  $n_{\rm GJ}\sim 5\times
10^{11}$ cm$^{-3}$,  often adopted as a reasonable estimate for the
charge density in the NS magnetosphere. Therefore, this hypothesis 
does not look very plausible.

Alternatively, one can assume that the spectral features are associated
with ion cyclotron energies, $E_{ci}=0.63 (Z/A) B_{14}$ keV,
where $Z$ and $A$ are the ion's charge and mass numbers.
The surface magnetic field needed for this interpretation
is $\ga 10^{14}$ G, much
larger than the conventional magnetic field inferred for a centered
dipole from the $P$, $\dot{P}$ measurements. 
However,  if the magnetic
dipole  is  off-centered to about 3 km below the surface, the 
magnetic moment $\mu = 3\times 10^{30}$ G cm$^3$ gives a surface
field of about $10^{14}$ G. We cannot exclude a possibility that
the true period derivative is significantly larger than that
estimated from the two observations (e.g., due to strong glitches),
which would increase the inferred magnetic field.  
However, the ion-cyclotron interpretation of the observed features
is not straightforward even in such a strong magnetic field.
An explanation of two features as the fundamental
and first harmonic of  an ion cyclotron line is unrealistic because the
ratio of the harmonic and fundamental  oscillator strengths,
$\sim E_{ci}/m_ic^2$, is extremely low. One could explain the two lines as
produced by different ions, with $(Z_1A_2)/(Z_2A_1)=2$. 
For instance, one might assume that the 1.4 keV and 0.7 keV lines are
due to protons ($Z/A=1$) and alpha-particles ($Z/A=1/2$), respectively,
in a magnetic field $B_{14}=2.2\, (1+z)$.
The problem with this
interpretation is that the strengths of the lines require similar
abundances of hydrogen and helium, which is hard to sustain because
of element sedimentation in the strong gravitational field of the NS.
To avoid this problem, one could assume that the radiation emerges from
a purely helium, partially ionized atmosphere --- then the 1.4 keV line
would be due to helium nuclei in $B_{14}=4.4\, (1+z)$,
while the 0.7
keV line would be caused by ion cyclotron transitions of once-ionized
helium ions ($Z/A=1/4$).  The required field, however, is so strong that
it is hard to expect a significant fraction of fully ionized helium at the
effective temperature 
of about 3 MK
inferred from the spectral continuum
(the temperature may be even lower if one applies
NS atmosphere models).
In addition, such cyclotron transitions
should be accompanied by atomic transitions in, at least, once-ionized helium.

\subsection{Atomic lines}

The other possibility is that the observed features are  atomic lines
formed in the NS atmosphere.
The available NS atmosphere models without magnetic field do not fit the
observed spectrum,
which is not surprising because the observed pulsations require a strong
magnetic field. 
In recent years there has been significant work on 
structure and spectra of atoms in strong magnetic fields, mostly in fields
below $10^{13}$ G (e.g., Ruder et al.\ 1994;  Mori \& Hailey 2002, and
references therein).  Based on these works, we can
exclude
some possibilities.  For instance, the observed absorption features
cannot be explained as emerging from a hydrogen atmosphere 
because, at any magnetic field and any reasonable gravitational
redshift, there is no pair of strong hydrogen
spectral lines whose energies would match the observed ones.
Therefore, one has to invoke heavier elements.
A possible interpretation of the observed features
as due to atomic transitions  of once-ionized helium
in a strong magnetic field,
$B\approx 1.5\times 10^{14}$ G, has been suggested by Pavlov et al.\ (2002b);
detailed modeling of the spectrum
will be presented in a forthcoming paper.
This interpretation yields the gravitational redshift $z=0.12$--0.23,
which corresponds to $R/M=8.8$--14.2 km~$M_\odot^{-1}$.
In that interpretation the 0.7 keV feature is a single line while
the 1.4 keV feature may include several lines which cannot be resolved
with ACIS.
To confirm this, one should investigate the detailed
structure of the detected strong features with a high-resolution
spectrometer and detect
other (weaker) helium lines.

\section{Conclusion}

The discovery of the absorption features in the spectrum of \ns\/ provides
the first opportunity to measure the mass-to-radius ratio and the
magnetic field of an isolated NS.  Measuring $M/R$ is particularly
important because it can constrain the equation of state of the
superdense matter in the NS interiors, infer the internal composition
of NSs, and test the theories of nuclear interactions.  Our analysis of
the low-resolution spectra has shown that interpreting the observed
features as cyclotron lines requires artificial assumptions, and such
features cannot be explained as formed in a hydrogen atmosphere.
The energies of the features suggest that they
could be
associated with the atomic transitions
of once-ionized helium in an atmosphere with a strong magnetic field.
This interpretation yields a gravitational redshift of about 0.17.
To firmly identify the features and measure the gravitational
redshift with better accuracy, deep observations with high spectral resolution are
needed, such that would be able to resolve potentially multiple
lines blended together due to the low spectral resolution
of the CCD detector.

\acknowledgements
We are grateful to Gordon Garmire, Leisa Townsley and George Chartas
for the useful advice on the analysis of ACIS data.  This work was partly
supported by SAO grant GO2-3088X and NASA grant NAG5-10865.
This research has made use of data obtained from the High Energy
Astrophysics Science Archive Research Center (HEASARC), provided
by NASA's Goddard Space Flight Center.

{}

\newpage

\begin{deluxetable}{cccccccccc}
\tabletypesize{\footnotesize}
\tablecolumns{10}
\tablewidth{0pc}
\tablecaption{
Fitting parameters for three models for two observations of \ns.}
\tablehead{
\colhead{} & \colhead{$n_{\rm H,20}$} & \colhead{$E_1$} & \colhead{$\tau_1$} &
\colhead{$\Gamma_1$} & \colhead{$W_1$} & \colhead{$E_2$} & \colhead{$\tau_2$} &
\colhead{$\Gamma_2$} & \colhead{$W_2$} \\ \\

\colhead{} & \colhead{} & \colhead{eV} & \colhead{} & \colhead{eV} &
\colhead{eV} & \colhead{eV} & \colhead{} & \colhead{eV} & \colhead{eV} \\
}

\startdata
\multicolumn{10}{c}{January 2000}\\ \\
1 &$2.33\pm0.44$ & $750\pm10$ & $0.39\pm0.03$ & $125\pm13$ & $108\pm12$
 & $1432\pm10$ & $0.45\pm0.04$ & $105\pm12$ & $102\pm14$ \\ \\
2 &$2.28\pm0.45$ & $749\pm10$ & $0.41\pm0.03$ & $134\pm14$ & $112\pm14$
 & $1432\pm10$ & $0.46\pm0.04$ & $112\pm14$ & $104\pm13$ \\ \\
3 &$2.34\pm0.44$ & $727\pm12$ & $0.40\pm0.06$ & $127\pm22$& $108\pm15$
 & $1401\pm12$ & $0.43\pm0.06$ & $112\pm24$ & $103\pm13$ \\ \\
\cline{1-10} \\
\multicolumn{10}{c}{January 2002} \\ \\
1 &$7.55\pm0.54$ & $735\pm14$ & $0.29\pm0.03$ & $94\pm20$ & $64\pm11$
 & $1423\pm10$ & $0.49\pm0.04$ & $125\pm14$ & $129\pm14$ \\ \\
2 &$7.49\pm0.54$ & $734\pm14$ & $0.30\pm0.03$ & $100\pm21$ & $68\pm13$
 & $1422\pm10$ & $0.50\pm0.04$ & $134\pm16$ & $130\pm16$ \\ \\
3 &$7.48\pm0.54$ & $721\pm17$ & $0.30\pm0.05$ & $98\pm36$ & $66\pm14$
 & $1380\pm13$ & $0.47\pm0.07$ & $131\pm26$ & $129\pm18$ \\ \\
\enddata

\tablecomments{
The ``optical depth'' $\tau_i$ characterizes the line flux with respect to the
continuum at the line center; for the second and third models it is defined
as $\exp(-\tau_i)\equiv r_i$ (see eqs.\ [2] and [3]). 
The errors quoted are the formal 1$\sigma$ uncertainties of
the fits.  The reduced $\chi_{\nu}^2$ values are within the range
1.03--1.06 (232 d.o.f. for the first observation and 236 d.o.f.
for the second observation).}
\end{deluxetable}

\newpage

\begin{figure}
\epsscale{0.85}
\plotone{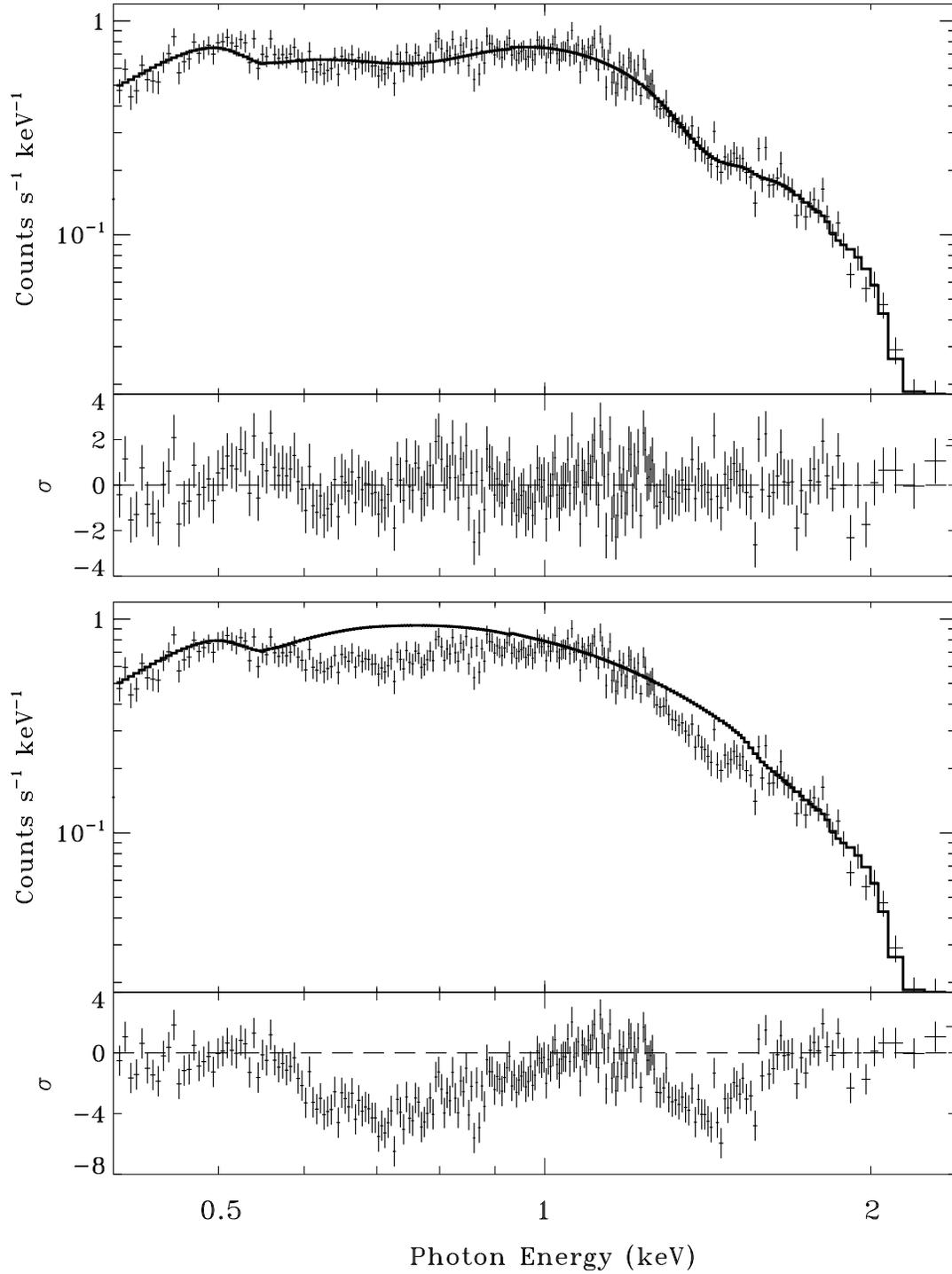}
\caption{The observed count spectrum and the best fit with model
(1) for the January 2000 observation (top panel).
The bottom panel shows the data and the continuum model.
\label{fig1}}
\end{figure}

\end{document}